\documentclass[aps,pra, preprint,groupedaddress,floatfix,longbibliography]{revtex4-1}
\usepackage{dcolumn}
\usepackage{graphicx}
\usepackage{amsmath}
\usepackage{bm}

\begin{document}


\title{Output Field-Quadrature Measurements and Squeezing in Ultrastrong Cavity-QED}


\author{Roberto Stassi$^{1,4}$}
\author{Salvatore Savasta$^{2,4}$}
\author{Luigi Garziano$^{2,4}$}
\author{Bernardo Spagnolo$^{1,3}$}
\author{Franco Nori$^{4,5}$}

\affiliation{$^1$Dipartimento di Fisica e Chimica, Group of Interdisciplinary Theoretical Physics, Universit\`{a} di Palermo and CNISM, Viale delle Scienze, I-90128 Palermo, Italy}
\affiliation{$^2$Dipartimento di Fisica e di Scienze della Terra, Universit\`{a} di Messina, Viale F. Stagno d'Alcontres 31, I-98166 Messina, Italy}
\affiliation{$^3$Istituto Nazionale di Fisica Nucleare, Sezione di Catania, via S. Sofia 64, I-95123 Catania, Italy}
\affiliation{$^4$CEMS, RIKEN, Saitama 351-0198, Japan}
\affiliation{$^5$Physics Department, The University of Michigan, Ann Arbor, Michigan 48109-1040, USA}


\date{\today}

\begin{abstract}
We study the squeezing of output quadratures of an electro-magnetic field escaping from a resonator coupled to a general quantum system with arbitrary interaction strengths. 
The generalized theoretical analysis of output squeezing proposed here is valid for all the interaction regimes of cavity-quantum electrodynamics: from the weak to the strong, ultrastrong, and deep coupling regimes. For coupling 
rates comparable or larger then the cavity resonance frequency, the standard input-output theory for optical cavities fails to calculate the correct output field-quadratures
and predicts a non-negligible amount of output squeezing, even if the system is in its ground state. Here we show that, for arbitrary interactions and cavity-embedded 
quantum systems, no squeezing can be found in the output-field quadratures if the system is in its ground state. We also apply the proposed theoretical 
approach to study the output squeezing produced by: (i) an artificial two-level atom embedded in a coherently-excited cavity; and (ii) a cascade-type three-level 
system interacting with a cavity field mode. In the latter case the output squeezing arises from the virtual photons of the atom-cavity dressed states. 
This work extends the possibility of predicting and  analyzing continuous-variable optical quantum-state tomography when optical resonators interact very strongly with other quantum systems.
\end{abstract}


\maketitle

\section{Introduction}

Recently, a new regime of cavity quantum electrodynamics (QED) has been experimentally reached in different solid state systems 
and spectral ranges~\cite{Forn-Diaz2010,Niemczyk2010, Todorov2010,Schwartz2011,Scalari2012,Geiser2012, Kena-Cohen2013,Gambino2014}. 
In this so-called ultrastrong coupling (USC) regime, where the light-matter coupling rate becomes an appreciable fraction of the 
unperturbed resonance frequency of the system, the routinely invoked rotating wave approximation (RWA) is no longer applicable 
and the antiresonant terms significantly change the standard cavity-QED 
scenarios~\cite{Dimer2007,DeLiberatoprl2007, Cao2010, Cao2011, Ridolfoprl2012, Ridolfoprl2013, Stassiprl2013, Sanchez-Burillo2014, Garziano2014, Cacciola2014}.

 
It has been shown that, in this USC regime, the correct description of the output photon flux, as well as of higher-order Glauber's normal-order correlation functions, requires a proper generalization of the input-output theory for resonators \cite{Ridolfoprl2012}.
The Application of the standard input-output picture to the USC regime would predict an unphysical continuous stream of output photons for a system 
in its ground state $| G \rangle$. This result stems from the finite number of photons which are present in the ground state due to 
the counter-rotating terms in the interaction Hamiltonian \cite{Ashhab2010}. Specifically, it has been shown~\cite{Ridolfoprl2012, Garziano2013} that the photon rate emitted by a resonator and
detectable by a photo-absorber is no longer proportional to $\langle  \hat a^\dag(t) \hat a(t) \rangle$ (as predicted by the standard input-output theory), 
where $\hat a$ and $\hat a^{\dagger}$ are the photon destruction and creation operators of the cavity mode, but to $\langle \hat x^-(t) \hat x^+(t) \rangle$, 
where  $\hat x^+(t)$ is the positive frequency component of the quadrature operator $\hat x(t) = \hat a(t) + \hat a^\dag(t)$ and $\hat x^-(t)=(\hat x^+(t))^\dagger$. 
This result shows that the excitations which are present in the ground state $|  G \rangle$, determining $\langle G| \hat a^\dag(t) \hat a(t)| G \rangle \neq 0$, actually do not correspond to physical observable particles. Only when the coupling rate is much smaller than the transition energies of the bare subsystems, then $\hat x^+ \to \hat a$. This situation displays interesting connections with quantum field theory 
where the creation and annihilation operators, present, e.g., in the Hamiltonian and Lagrangian of QED, describe the creation and 
destruction of bare particles which, however, cannot be directly observed in experiments. Physical particles in quantum field 
theory, due to the interaction terms in the Lagrangian, are actually surrounded by clouds of virtual particles. Moreover, as in quantum field theory, in the USC regime, the total
number of excitations in the cavity-emitter system is not preserved.

Direct photon counting experiments provide information about the mean photon number and higher-order normal-order correlations. However a complete quantum tomography of the electro-magnetic field (see, e.g., \cite{Lvovsky2009}) requires phase-sensitive measurements which are based on homodyne or  
heterodyne detection \cite{Wiseman1993,Mallet2011}. These techniques enable the measurements of the mean field quadratures and their variance, e.g., $\langle \hat x\rangle$ and $\langle \hat x^2\rangle - \langle \hat x\rangle^2$.

In a coherent state of an  electro-magnetic field mode,  the quantum fluctuations of  two field-quadratures $\hat Q_1$ and $\hat Q_2$  with $ [ \hat Q_1,\hat Q_2]  =1$  are equal
($\Delta \hat Q_1=\Delta \hat Q_2=1$ where $\Delta \hat Q_i =\langle \hat Q_i^2\rangle - \langle \hat Q_i\rangle^2$) and minimize the uncertainty product given by Heisenberg's uncertainty relation $\Delta \hat Q_1\, \Delta \hat Q_2= 1$ 
(we use $\hbar=1$). These zero-point fluctuations represent the standard quantum limit to the reduction of noise in a signal. 
Other minimum uncertainty states are possible, and these occur when  fluctuations in one quadrature are squeezed at the expense of increased fluctuations in the other one \cite{Drummond2013}. Light 
squeezing can be realized in various nonlinear optical processes, such as parametric down-conversion, parametric amplification, and degenerate four-wave mixing \cite{Walls2007,Mandel1995,Scully1997, Bartkowiak2014}  or in presence of time-dependent boundary conditions \cite{Zagoskin2008,Johansson2010, Wilson2011,Norirmp}.
Squeezed states of light belong to the class of nonclassical states of light.
Having a less noisy quadrature, squeezed light has applications in optical communication \cite{Slavik2010} and measurements \cite{Caves1981, Braunstein2005, Castellanos-Beltran2008,Slavik2010, Giovannetti2011} and is a primary resource in continuous variable
quantum information processing\cite{ Braunstein2005}. 
Squeezing of the electromagnetic field has been achieved in a variety of systems operating in the optical and microwave regimes. 
A noise reduction of -10 dB (-13 dB is the estimation of squeezing after correction for detector inefficiency)  achieved in the experiment~\cite{Vahlbruch2008}. More recently, a
few experiments with superconducting circuits~\cite{Wilson2011,Eichler2011} have
demonstrated the possibility of obtaining much stronger squeezing in
microwave fields~\cite{Flurin2012}.

%
%
%
%
In the atom-cavity coupled system, the squeezing effect has been usually studied by using the rotating-wave approximation~\cite{Walls1981,Meystre1982,carmichael86, kimble87, nha03}.

Here we present a theory of  quadrature measurements of the output field escaping from a resonator coupled to a generic matter 
system with arbitrary interaction strength, and we apply it to the analysis of squeezing. While in the ultrastrong coupling regime the positive frequency 
component $\hat x^+$ is different from $\hat a$, the quadrature operator $\hat x= \hat a + \hat a^\dag = \hat x^+ + \hat x^-$ is independent of the light-matter interaction 
strength. Hence, at a first sight, one may expect that, in contrast to Glauber's correlation functions, quadrature measurements can be analyzed 
by applying the standard input-output theory \cite{Collett1984,Gardiner1985}. Here we show that this is not the case. Application of the standard input-output 
picture to the analysis of quadrature measurements in the ultrastrong coupling regime leads to incorrect results.

We apply the theoretical framework here developed to the case of a coherently-excited cavity
interacting with an artificial two-level atom. We also analyze the output  squeezing  from a resonator interacting with a cascade-type three-level system. In the latter case, the output squeezing arises from the virtual photons carried by the atom-cavity dressed  states.
We also analyze the output field-quadratures  for a system in its ground state. It is know that the 
ground state of a system in the ultrastrong coupling regime is a squeezed vacuum state, where the amount of squeezing depends on the coupling 
strength and on the detuning between the cavity mode and the matter system resonances. Recently, a correlation-function analysis of 
the quadratures of microwave fields has been exploited for measurements of vacuum fluctuations and weak thermal fields \cite{Mariantoniprl2010}. Hence the 
question arises if it is possible to detect such vacuum squeezing.

In  harmonic interacting systems described by time-independent Hamiltonians, it has been shown that it is not possible to detect any squeezing from a system in the vacuum state~\cite{Savasta1996,Ciuti2006} or under coherent driving . This result can be 
understood considering that the physical quanta that can be detected are  polaritons, hybrid Bosonic particles which do not display any squeezing, and not the bare photons which can display squeezing. Of course, it is not surprising that, though strongly interacting, harmonic systems which can be described just as a 
collection of noninteracting harmonic oscillators do not display squeezing in the vacuum state or when coherently excited. Generally 
speaking, the appearance of detectable nonclassical correlations requires systems subject to some degree of nonlinearity or a time-dependent Hamiltonian. The simple system composed by a qubit coupled to a resonator in the ultrastrong coupling regime is highly 
anharmonic, hence we may ask if its ground state can give rise to any detectable squeezing in the output field. Here we demonstrate 
that for arbitrary cavity-embedded quantum systems, independently on the coupling rate, no squeezing can be found in the output field 
quadratures if the system is in its ground state.


\section{Squeezing of the ground state of the Rabi Hamiltonian}

The Hamiltonian of the quantum Rabi model ($\hbar =1$) \cite{Rabi1936,Rabi1937}
is given by
\begin{equation} \label{HR}
\hat H_{\rm R} =\omega_{\rm c}\, \hat a^\dagger \hat a +\frac{\omega_{\rm q}}{2} \hat \sigma_z + \Omega_{\rm R} \left(\hat a^\dagger +\hat a \right) \hat \sigma_x\, ,
\end{equation}
where $\hat a$ and $\hat a^\dag$ are, respectively, the annihilation and creation
operators for the cavity field of frequency $\omega_{\rm c}$. The Pauli
matrices are defined as $\hat \sigma_z = |e \rangle \langle e| - |g \rangle \langle g|$ and $\hat \sigma_x = \hat \sigma_+ + \hat \sigma_- = |e \rangle \langle g| + |g \rangle \langle e|$, in terms of the atomic ground ($|g \rangle$) and excited ($|e \rangle$) states. The parameter $\omega_{\rm q}$ describes the transition energy of the two-level system  and $\Omega_{\rm R}$ is the coupling energy between the atomic transition and the cavity field. 

Owing to the presence of the so-called counter-rotating terms,  $\hat a \hat \sigma_-$ and  $\hat a^\dag \hat \sigma_+$, in the Rabi Hamiltonian, the operator describing the total number of excitations, $\hat N = \hat a^\dag \hat a + |e \rangle \langle e|$, does not commutes with $\hat H_{\rm R}$ and as a consequence the eigenstates of $\hat H_{\rm R}$ do not have a definite number of excitations \cite{Ashhab2010}. For instance the resulting ground state is a superposition of an even number of excitations, 
\begin{equation}
| G \rangle  \equiv |\tilde 0 \rangle = \sum_{k= 0}^\infty (c^{\tilde 0}_{g,2 k} |g, 2 k \rangle 
+ c^{\tilde 0}_{e, k} |e,  2k+1 \rangle)\, ,
\label{zerotilde}
\end{equation}
where the second entry in the kets provides the photon number (see, e.g., Refs.\, \cite{Ashhab2010, Garziano2013}).
When the coupling rate $\Omega_{\rm R}$ is much smaller than the bare resonance frequencies of the two subsystems  $\omega_{\rm c}$ and $\omega_{\rm q}$, only $c^{\tilde 0}_{g,0}$ is significantly different from zero and the ground state reduces to $| \tilde 0 \rangle \simeq| g,0 \rangle$, which is that of the Jaynes-Cummings model, derived from the Rabi Hamiltonian after dropping the counter-rotating terms.
When the coupling rate $\Omega_{\rm R}$ approaches and exceed $10 \%$ of the bare frequencies of the subsystems (ultrastrong coupling regime), contributions with $k \neq 0$ in Eq.\ (\ref{zerotilde}) become not negligible. One consequence is that the mean photon number in the ground state  $\langle \tilde 0 | \hat a^\dag \hat a | \tilde 0 \rangle$ becomes different from zero. Moreover, the ground state displays a certain amount of photon squeezing.
Considering the intracavity-field quadrature $\hat q_2 =i (\hat a^\dag - \hat a)$, its variance $s_2 = \langle \tilde 0| \hat q_2^2 |\tilde 0 \rangle -  \langle \tilde 0| \hat q_2 |\tilde 0 \rangle^2$ (notice that $\langle \tilde 0| \hat q_2 |\tilde 0 \rangle = 0$) turns out to be below the standard quantum limit value $1$.
\begin{figure}[!ht]
	\includegraphics[width=80 mm]{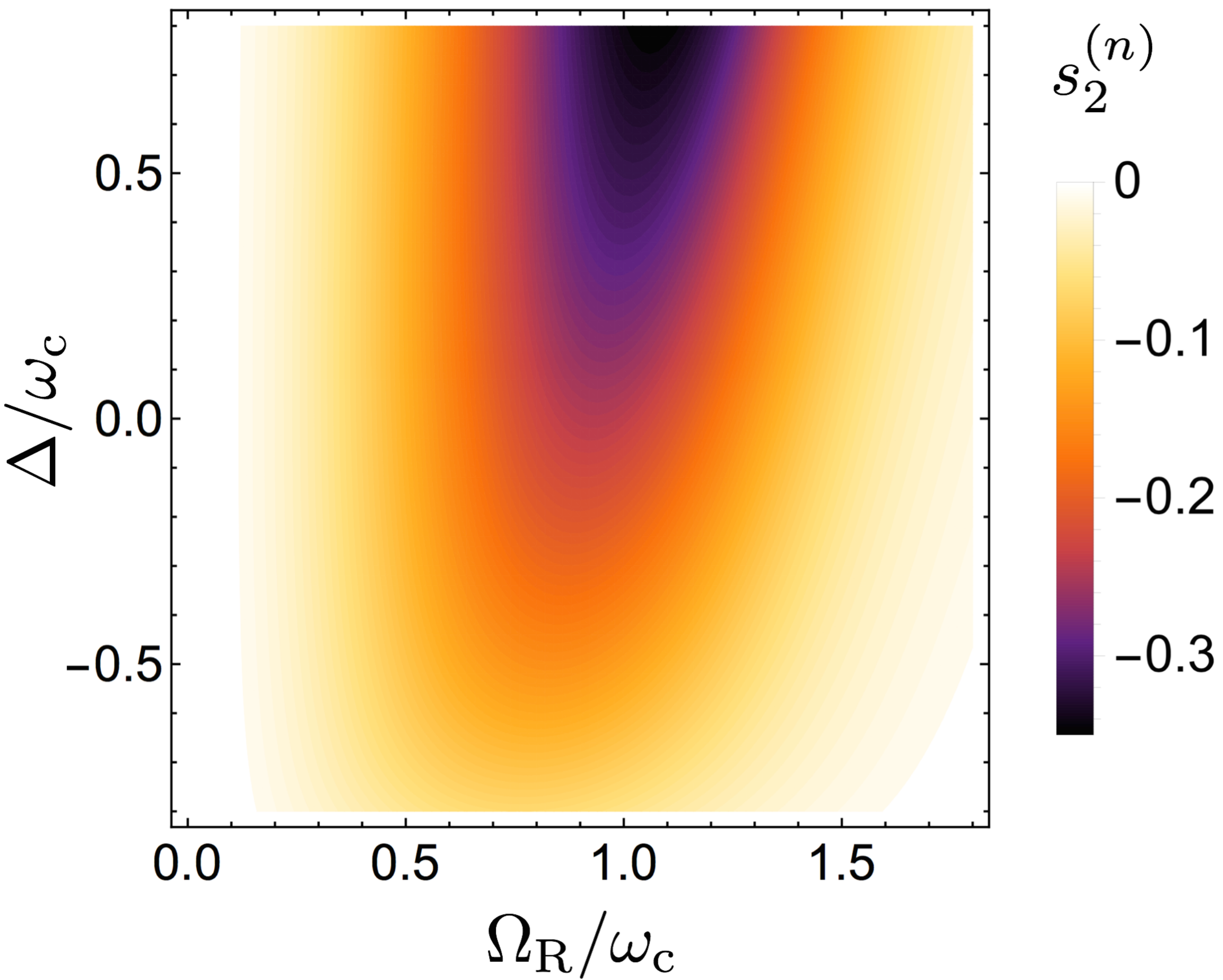}
	\caption{(Color online) Normally-ordered variance $s^{(n)}_2 = s_2 -1$ of the cavity-field quadrature $\hat x_2 =i (\hat a^\dag - \hat a)$, calculated in the ground state $| \tilde 0 \rangle$ of the Rabi Hamiltonian, as a function of the normalized coupling rate $\Omega_{\rm R}/\omega_c$ and cavity-atom detuning $\Delta/\omega_{\rm c}=(\omega_{\rm q}-\omega_{\rm c})/\omega_{\rm c}$.}\label{S2}
\end{figure}

Figure~1 displays the numerically calculated $s_2^{(n)}=s_2-1$ as a function of the normalized coupling $\Omega_{\rm R}/\omega_{\rm c}$ and  detuning $\Delta/\omega_{\rm c}=(\omega_{\rm q}-\omega_{\rm c})/\omega_{\rm c}$. For small values of the normalized coupling, the variance approaches the standard quantum limit. Increasing $\Omega_{\rm R}/\omega_{\rm c}$, the variance decreases below the standard quantum limit, reaching a lowest value of about $-0.35$, at $\Omega_{\rm R}/\omega_{\rm c} \simeq 1.05$ and at a positive detuning $\Delta/\omega_{\rm c} \simeq 0.76$.
Further increasing the coupling, expecially at zero and negative detuning, results into an increase of the variance $s_2$, caused by quite large contributions in $|\tilde 0 \rangle$ of terms with an odd number of photons.

In the next section we will show that such a ground-state squeezing does not give rise to an observable output squeezing.

\section{Output field quadratures}
According to the input-output theory for general localized quantum systems interacting with a propagating quantum field, the output field 
operator can be related through a boundary condition to a system operator and the input field operators~\cite{Gardiner2004}. In order 
to be specific, we consider the case of a system coupled to a semi-infinite transmission line~\cite{Gardiner2004}, although the 
results obtained can be applied or extended to a large class of systems. While the resonator can be ultrastrongly coupled to a 
localized quantum system, its interaction with the propagating quantum field (e.g., the transmission line) is weak. To derive the \textit{input-output} 
relations we couple the system to a quantum field made of an assembly of harmonic oscillators. The total Hamiltonian of the system 
can be written  as
\begin{equation}
  \hat H=\hat H_{\rm S}+\hat H_{\rm F}+\hat H_{\rm SF},
\end{equation}
where $\hat H_{\rm S}$ and $\hat H_{\rm F}$ are the system and field Hamiltonian and where the interaction between the system and the field can be expressed in the rotating wave approximation as
\begin{equation}\label{HSB}
  \hat H_{\rm SF} = i \int_0^\infty  d\omega\, k\left(\omega\right)\sqrt{\frac{v \hbar\omega}{2}} \left[ \hat X^- \hat b\left(\omega \right)-\hat X^+ 
  \hat b^\dagger\left(\omega \right)\right]\, ,
\end{equation}
where $v$ is the speed of the travelling field, e.g., the speed of light in the transmission line.
In the above equation, $\hat b\left(\omega \right)$ is the annihilation operator for the harmonic oscillators that describe the field, $\hat X^+$ and $\hat X^-$ 
are the \textit{positive} and \textit{negative} frequency components of the generic system operator $\hat X$ coupled to the field. These components 
can be obtained expressing $\hat X$ in the eigenvectors basis of $\hat H_{\rm S}$ as $\hat X^+=\sum_{i<j}X_{ij}\left|i\rangle\langle j\right | $, 
and $\hat X^-=(\hat X^+)^\dagger$.  Here  the eigenstates of $\hat H_{\rm S}$ are labeled according to their eigenvalues such that  $\omega_{k} > \omega_{j}$ for $k>j$. 
We observe that the rotating wave approximation used in Eq.\ (\ref{HSB}) is based on the separation into positive and negative frequency operators of the system operator $\hat X$ after the system diagonalization. The standard RWA is instead  based on the separation into bare positive (destruction) and negative (creation) components of the field operator coupled to the external modes,  without including its interaction with other components of the system.

The \textit{positive} frequency component of the input and output fields can be written as
\begin{equation}\label{Ain}
\hat A_{\rm in\left(\rm out\right)}^{+}\left(t \right)=\frac{1}{2}\int_{0}^{\infty}\mathrm{d}\omega\sqrt{\frac{\hbar}{\pi\omega }}\hat b\left( \omega,t'\right) 
e^{-i \omega\left( t-t'\right)}\, ,
\end{equation}
while the \textit{negative} frequency component $\hat A_{\rm in\left(\rm out\right)}^{-}=\left(\hat A_{\rm in\left(\rm out\right)}^{+}\right)^\dagger$, 
so that $\hat A_{\rm in\left(\rm out\right)}\left(t \right)=\hat A_{\rm in\left(\rm out\right)}^{+}\left(t \right)+\hat A_{\rm in\left(\rm out\right)}^{-}\left(t \right)$, 
in which $t'<t$ (the input) is an initial time and $t'>t$ (the output) is assumed to be in the remote future\,\cite{Gardiner2004}. Formally 
solving the Heisenberg equations of motion for $\hat b(\omega)$, the input-output relations for the positive and negative components of the 
fields can be obtained \cite{Garziano2013}
\begin{equation}\label{inout}
\hat A_{\rm out}^{\pm}\left(t \right)=\hat A_{\rm in}^{\pm}\left(t \right)-\gamma \hat X^\pm\left(t \right)\, ,
\end{equation}
where for the sake of simplicity  the first Markov approximation, $k(\omega)=\sqrt{2\gamma/\pi}$, has been adopted. However, the present 
analysis can be easily extended beyond these approximation. Equation (\ref{inout}) shows that the positive frequency  output operator can be 
expressed in terms of the positive frequency input operator and the positive frequency system operator coupled to the propagating field. 
If the system consists of an empty single-mode resonator, then $\hat X^+ \propto \hat a$, being $\hat a$ the destruction operator of the cavity mode. If instead 
the cavity mode is coupled to another quantum system, e.g., an atom, $\hat X^+$ will be different from $\hat a$, and may also contain contributions 
from $\hat a^\dag$. In this case, the positive component of the output field may contain contributions from the creation cavity operators, in contrast to 
ordinary quantum optical input-output relationships \cite{Gardiner1985,Walls2007}.

We define the output quadrature operators $\hat Q_{1}\left(t \right)$ and $\hat Q_{2}\left(t \right)$ as
\begin{equation}\label{X1out}
\begin{split}
& \hat Q_{1}\left(t\right)=\hat A_{\rm out}^{+}\left(t \right)e^{-i \Gamma(t)}+\hat A_{\rm out}^{-}\left(t \right)e^{i\Gamma(t)}\\
& \hat Q_{2}\left(t\right)=-i \left[\hat A_{\rm out}^{+}\left(t \right)e^{-i\Gamma(t)}-\hat A_{\rm out}^{-}\left(t \right)e^{i\Gamma(t)}\right]
\end{split}
\end{equation}
so that
\begin{equation}\label{AoutX12}
\hat A_{\rm out}^{\pm}\left(t \right)=\frac{1}{2}\left[\hat Q_{1}\left(t \right) \pm i \hat Q_{2}\left(t \right)   \right]e^{i\Gamma(t)}
\end{equation}
where $\Gamma(t)=\Omega t+\varphi$, $\Omega$ and $\varphi$ are the reference frequency and phase, respectively. It is possible to pass from one 
quadrature to the other applying a $\pi/2$ rotation.

By solving the Heisenberg equations for the field operators $\hat b(\omega)$ and using Eq.~(\ref{Ain}), we  obtain the following commutation relation
between any system variable $\hat Y(t)$ and  the input fields $\hat A^\pm(t)
$\begin{equation}\label{com}
\left[\hat Y\left(t \right),\hat A_{\rm in}^{\pm}\left(s \right)\right]=\sqrt{\gamma}\,u\left(t-s \right)\left[\hat Y\left(t \right),\hat X^\pm\left(s \right)\right]\, ,
\end{equation}
where $u\left(t-s \right)$ is egual to $1$ if $t>s$, $\frac{1}{2}$ if $t=s$, $0$ if $t<s$.
Making use of the \textit{input-output} relations (\ref{inout}) and  of the commutation relations (\ref{com}) we can proceed to calculate 
the output field quadrature variances $S_i\left(t,\tau \right)=\langle \hat Q_i \left(t \right), \hat Q_i \left(t+\tau \right) \rangle$ in 
terms of correlation functions involving only input operators or system operators (here we used $\langle \hat A, \hat B \rangle = \langle \hat A \hat B \rangle - \langle \hat A \rangle \langle \hat B \rangle$).
Considering an input in a vacuum or a coherent state, the field-quadrature variances can be expressed as
\begin{eqnarray}\label{squeezingtau}
S_1\left(t,\tau\right) &=&\gamma\, \left[{\cal T}\langle \hat X^+(t+\tau),\hat X^+(t) \rangle e^{-2i\Gamma(t)} + {\cal T}\langle \hat X^-(t),\hat X^-(t+\tau) \rangle e^{2i\Gamma(t)}+ \langle \hat X^-(t+\tau),\hat X^+(t)\rangle \right. \nonumber \\ 
&&+\left. \langle \hat X^-(t),\hat X^+(t+\tau)\rangle \right]+ \langle \hat A^+_{\rm in}(t),\hat A_{\rm in}^-(t+\tau)\rangle\, ,
\end{eqnarray}
where ${\cal T}$ is the time-ordering operator such that rearranges creation operators in forward time and annihilation operators in backward time order.
 To obtain $S_2$ we can apply a $\pi/2$ rotation to Eq.\,(\ref{squeezingtau}) substituting  $\theta\to\theta+\pi/2$. For equal-time correlation functions  ($\tau=0$), we have
\begin{eqnarray}\label{squeezing}
S_1\left(t \right) &=&\gamma \left[\langle \hat X^+(t),\hat X^+(t) \rangle e^{-2i\Gamma(t)} + \langle \hat X^-(t),\hat X^-(t) \rangle e^{2i\Gamma(t)}+2 \langle \hat X^-(t),\hat X^+
(t)\rangle \right] \nonumber \\ &&+ \langle \hat A^+_{\rm in}(t),\hat A_{\rm in}^-(t)\rangle\, .
\end{eqnarray}
The last term in Eq.\ (\ref{squeezing}) $\langle \hat A^+_{\rm in},\hat A_{\rm in}^-\rangle$ describes the quantum noise of the input port in the vacuum state. If in addition the system is in its ground state $|G \rangle$, $\hat X^+(t) |G \rangle =0$ (of course, in this case we consider a vacuum input), the output noise coincides with the input one, $S_1\left(t \right) = \langle \hat A^+_{\rm in}(t)\hat A_{\rm in}^-(t)\rangle$.

From Eq.~(\ref{squeezing}) we can formulate the following general statement: \textit{Any open system  in its ground state, i.e., $\hat X^+\left |0 \right>=0$, does not display any output squeezing  (even if its ground state is a squeezed state)}. 
 This absence of output ground-state squeezing has been previously shown in different interacting harmonic systems. 
Equation (\ref{squeezing}) holds for general open quantum systems, independently their composition in subsystems and the degree of interaction among the different subsystems. In order to compare this result with previous descriptions for optical resonators, we consider the case where $\hat X$ describes the field of a single-mode cavity: $\hat X = X_0 \hat x = X_0 (\hat a + \hat a^\dag)$. Here $X_0$ denotes the zero-point fluctuation amplitude of the resonator. Equation  (\ref{squeezing}) can be expressed as 
\begin{eqnarray}\label{squeezing2}
S_1\left(t \right) &=&\gamma  X_0^2 \left[\langle \hat x^+(t),\hat x^+(t) \rangle e^{-2i\Gamma(t)} + \langle \hat x^-(t),\hat x^-(t) \rangle e^{2i\Gamma(t)}+2 \langle \hat x^-(t),\hat x^+
(t)\rangle \right] \nonumber \\ &&+ \langle \hat A^+_{\rm in}(t),\hat A_{\rm in}^-(t)\rangle\, .
\end{eqnarray}
If the interaction of the resonator with other quantum systems is not in the USC regime, $\hat x^+ =\hat a$ and $\hat x^- = \hat a^\dag$. The noise reduction with respect to the vacuum input can be expressed in terms of the following normally-ordered variance
\begin{equation}
S^{(n)}_i\left(t \right) = \frac{S_i\left(t \right)- \langle \hat A^+_{\rm in}(t),\hat A_{\rm in}^-(t)\rangle}{\gamma X_0^2}
\end{equation}
For a resonator not in the USC regime, an ideally squeezed quadrature corresponds to $ S^{(n)}_i = -1$, while for a resonator in the ground state $S^{(n)}_i = 0$.


\section{Squeezing of output field-quadratures in the USC regime}

Here we apply the theoretical framework developed in Sect.\,III to study the output field-quadrature variances in single-atom USC cavity-QED systems.
We first consider the case of a flux qubit artificial atom  coupled to a  $\lambda/2$ superconducting transmission-line resonator, when the frequency of the resonator is near one-half of the atomic transition frequency (see Fig.\ 2). Recently it has been shown \cite{Garziano2015} that this regime can strongly modify the concept of vacuum Rabi oscillations, enabling two-photon exchanges between the qubit and the resonator. Here we show that such  configuration can provide a very large amount of squeezing although the system has only one artificial atom and displays a moderate coupling rate $\Omega_{\rm R}/\omega_{\rm c} \sim 0.1$. Then, we will study the output squeezing of a cascade three-level system where only the upper transition is coupled to the optical resonator.

 In order to describe a realistic system, the dissipation channels need to be taken into account. For this reason all the dynamical evolutions displayed below have been numerically calculated solving the  master equation $\dot{\hat \rho}(t) = i [\hat \rho(t), H] + \sum_{\rm i} \mathcal{L}_{\rm i}\hat \rho(t)$ \cite{Breuer2002, Blais2011, Garziano2013}, where $\mathcal{L}_{\rm i}$ is a Liouvillian superoperator describing the cavity and atomic system losses (see Appendix A). All calculations have been carried out by considering zero temperature reservoirs.

\subsection{Two-photon Rabi oscillations}

\begin{figure}[!ht]
	\includegraphics[width= 80 mm]{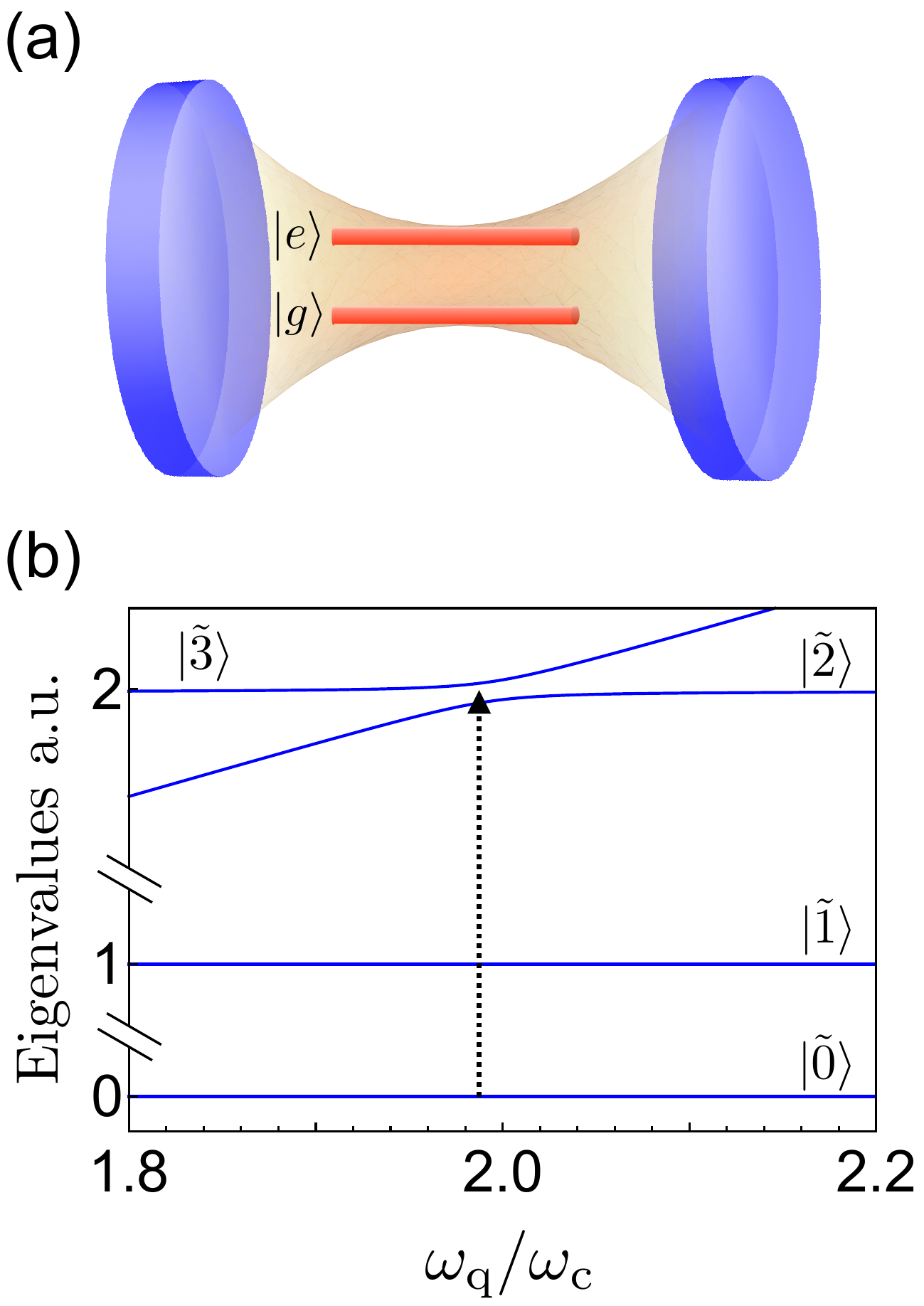}
	\caption{(a) Sketch of a cavity-embedded two-level system.
	(b) Frequency differences with respect to the ground state: $\omega_{\tilde k \tilde 0}=\omega_{\tilde k}-\omega_{\tilde 0}$ for the lowest-energy dressed states of $\hat H'_{\rm{R}}$ as a function of the qubit transition frequency $\omega_{\rm q}/\omega_{\rm c}$. We consider a normalized coupling rate $\Omega_{\rm R}/\omega_{\rm c}=0.15$ between the qubit and the resonator. In correspondence of the avoided level crossing (at $\omega_{\rm q} \approx 2 \omega_{\rm c}$) a Gaussian pulse is sent with central frequency in the middle of the two split transition energies (black arrow).}\label{lev}
\end{figure}
We now consider a flux qubit ultrastrongly coupled to a coplanar resonator \cite{Niemczyk2010} (see Fig.\,\ref{lev}).  In this  system both the number of excitations and parity symmetry are no longer conserved and transitions which are forbidden in natural atoms become available \cite{Liu2005}. This paves the way to anomalous vacuum Rabi oscillations, where two or more photons are jointly and reversibly  emitted and reabsorbed by the qubit \cite{Law2015,Garziano2015}. 

This quantum circuit can be described by the following extended Rabi Hamiltonian \cite{Niemczyk2010}
\begin{equation}
  \hat H'_{\rm{R}}=\omega_{\rm c}\, \hat a^\dagger \hat a +\omega_{\rm q}\hat \sigma^+\hat \sigma^-+\Omega_{\rm R}\left(\hat a^\dagger +\hat a \right)
 (\cos \theta\,  \hat \sigma_x + \sin \theta\, \hat \sigma_z ).
\end{equation}
The angle $\theta$ as well as the qubit resonance frequency depend on the  flux offset $\delta \Phi_{\rm q} \equiv \Phi_{\rm ext} - \Phi_0$, where $\Phi_{\rm ext}$ is the external magnetic flux threading the qubit and $\Phi_0$ is the flux quantum. A flux offset $\delta \Phi_{\rm q} = 0$ implies $\theta =0$. In this case $H'_{\rm R}$ reduces to the standard Rabi Hamiltonian (\ref{HR}).
We choose the labelling of the eigenstates  $|\tilde{i}\rangle$ and eigenvalues  $\omega_{\tilde j}$ of $H'_{\rm R}$ such that $\omega_{\tilde k} > \omega_{\tilde j}$ for $\tilde k > \tilde j$.

The lowest eigenenergy offsets with respect to the ground energy  $\omega_{\tilde j}- \omega_{\tilde 0}$ as a function of the qubit transition frequency $\omega_{\rm q}$ are shown in Fig.\ 2. Looking at the numerically calculated  eigenvectors, the first excited state, $|\tilde{1}\rangle$, contains a dominant contribution from the bare state $|g, 1\rangle$, ($|\tilde{1}\rangle\simeq |g, 1\rangle$). The figure also shows an avoided crossing when $\omega_{\rm q} \approx 2 \omega_{\rm c}$.
The splitting can be attributed to the resonant coupling of the states $|e,0 \rangle$ and $|g, 2 \rangle$, although the USC regime implies that the resulting dressed states $| \tilde 2 \rangle$ and $| \tilde 3 \rangle$ contain also small contributions from other bare states, as  $| g,1 \rangle$ and  $| e,1 \rangle$.
This splitting cannot be found in the rotating wave approximation, where the coherent coupling between states with a different number of excitations is not allowed, nor does it occur with the standard Rabi Hamiltonian ($\theta =0$).

We consider a system initially in the ground state. Excitation occurs  by direct optical driving of the qubit via a microwave antenna. The corresponding driving Hamiltonian is
\begin{equation}\label{pulse}
\hat H_{\rm d} = {\cal E}(t) \cos (\omega t) \hat \sigma_x\,,   
\end{equation}
where ${\cal E}(t) = A \exp{[-(t-t_0)^2/(2 \tau^2)]}/(\tau \sqrt{2 \pi})$ describes a  Gaussian pulse. Here $A$ and $\tau$ are the amplitude and the standard deviation of the Gaussian pulse, respectively. We consider the zero-detuning case, corresponding to the minimum energy splitting $2 \Omega_{\rm eff}$ in Fig.\,2b. The central frequency of the pulse has been chosen to be in the middle of the two split transition energies: $\omega = (\omega_{\tilde 3} + \omega_{\tilde 2})/2- \omega_{\tilde 0}$. 
If $\tau$ is much smaller than the effective Rabi period, $ \tau \ll T_{\rm R} = 2 \pi / \Omega_{\rm eff}$, the driving pulse is able to generate an initial  superposition with equal weights of the states $| \tilde 2 \rangle$ and $| \tilde 3 \rangle$, which will evolve displaying two-photon quantum vacuum oscillations \cite{Garziano2015}.
Figure 3a displays the resulting qubit population  (red dashed curve) and mean photon number (blue continuous) after a pulsed excitation with an effective pulse area ${\cal A} = \pi/3$. 
Figure 3b shows the normally ordered variance of the two orthogonal output field quadratures $S^{(n)}_1$ (blue continuous curve) and  $S^{(n)}_2$ (dotted red).
Both the two quadratures display a significant amount of squeezing when the mean photon number is maximum. It is interesting to see that the periodicity of the two variances is twice the Rabi period $T_{\rm R}$. This can be understood noticing that after the excitation, the quantum state is a superposition of the ground state $| \tilde 0 \rangle$ and the excited states  $| \tilde 2 \rangle$ and $| \tilde 3 \rangle$. After one Rabi oscillation, the excited states acquire a $\pi$ phase shift.
A second Rabi oscillation is needed to recover the initial phase.
The dynamics of the corresponding variances (not shown here) calculated by using $\hat a$ and $\hat a^\dag$, instead of $\hat x^+$ and $\hat x^-$, are affected by fast oscillations.

\begin{figure}[!ht]
	\includegraphics[width=100 mm]{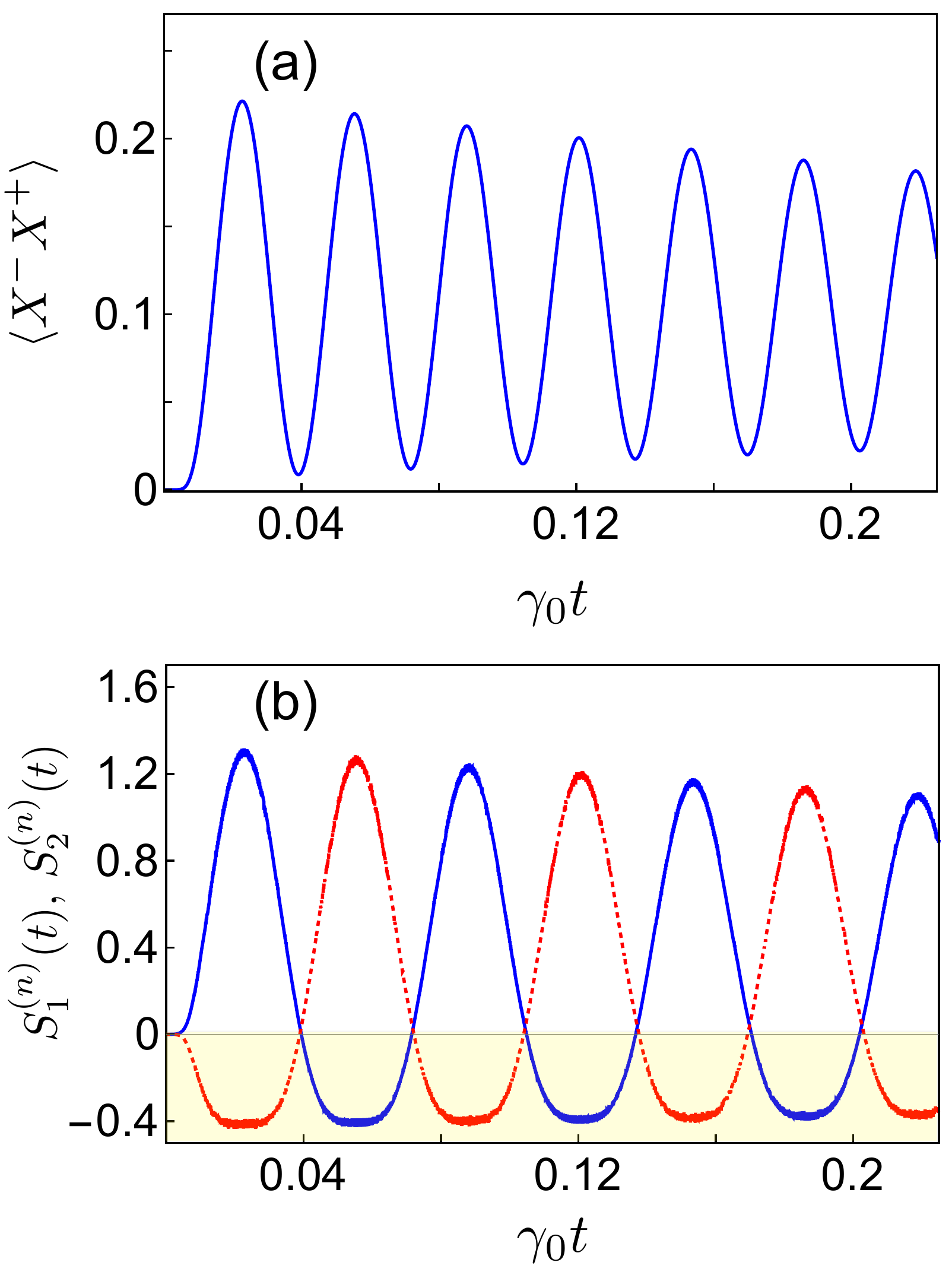}
	\caption{(Color online) (a) Temporal evolution of the cavity mean photon number $\langle \hat X^-\hat X^+\rangle$ (blue continuos curve) after the arrival of a Gaussian pulse exciting the qubit. The pulse has an affective area $\pi/3$ and central frequency $(\omega_{\tilde 3} + \omega_{\tilde 2})/2$. (b) Time evolution of the normally-ordered variances $S^{(n)}_1(t)$ (blue continuos curve) and $S^{(n)}_2(t)$ (red dashed curve). Here, the resonator and qubit damping rates are $\gamma_c=\gamma_q=1.8\times 10^{-4}\omega_{\rm c}$. The yellow background shows the region with squeezed states.}\label{2lev}
\end{figure}
This periodic and alternating squeezing of the two quadratures can be better understood by a simplified effective model assuming that 
\begin{eqnarray}\label{states1}
| \tilde 2 \rangle &\simeq& \frac{1}{\sqrt{2}}(|e,0 \rangle  + |g, 2 \rangle)\,, \nonumber \\
| \tilde 3 \rangle &\simeq& \frac{1}{\sqrt{2}}(|e,0 \rangle  - |g, 2 \rangle)\, .
\end{eqnarray}
Considering the qubit initially prepared in the superposition state $ |\psi(t=0) \rangle = \alpha |g,0 \rangle + \beta |e,0 \rangle$ (with $| \alpha|^2 + |\beta|^2 =1$), the resulting  time evolution of the system state is, to a good approximation,
\begin{equation}
 |\psi(t) \rangle = \alpha  |g,0 \rangle +  \beta \left[\cos{(\Omega_{\rm eff} t)} |e,0 \rangle + \sin (\Omega_{\rm eff} t) |g,2 \rangle \right]\, ,
\end{equation}
where $2 \Omega_{\rm eff}$ is the minimum energy splitting in Fig.\,2b.
 At $t= \pi/(2 \Omega)$, the resulting state is $ |g\rangle ( \alpha |0 \rangle +  \beta |2 \rangle )$, which is a squeezed photon state,  reaching a maximum squeezing for $\alpha \simeq 1/3$.

\subsection{Cascade three-level system}

We consider  a three-level $(|s\rangle, |g\rangle$ and $|e\rangle)$ atom-like system  with the upper transition $(|g\rangle\leftrightarrow |e\rangle)$ ultrastrongly coupled with a mode of the resonator and a lower transition which does not interact with the resonator, as schematically shown in Fig.\ 4. The peculiar optical properties of this system have been analyzed calculating the dynamics of the populations and of normal-order correlation functions \cite{Ridolfo2011,Stassiprl2013,Huang2014}.
\begin{figure}[!ht]
	\includegraphics[width=80 mm]{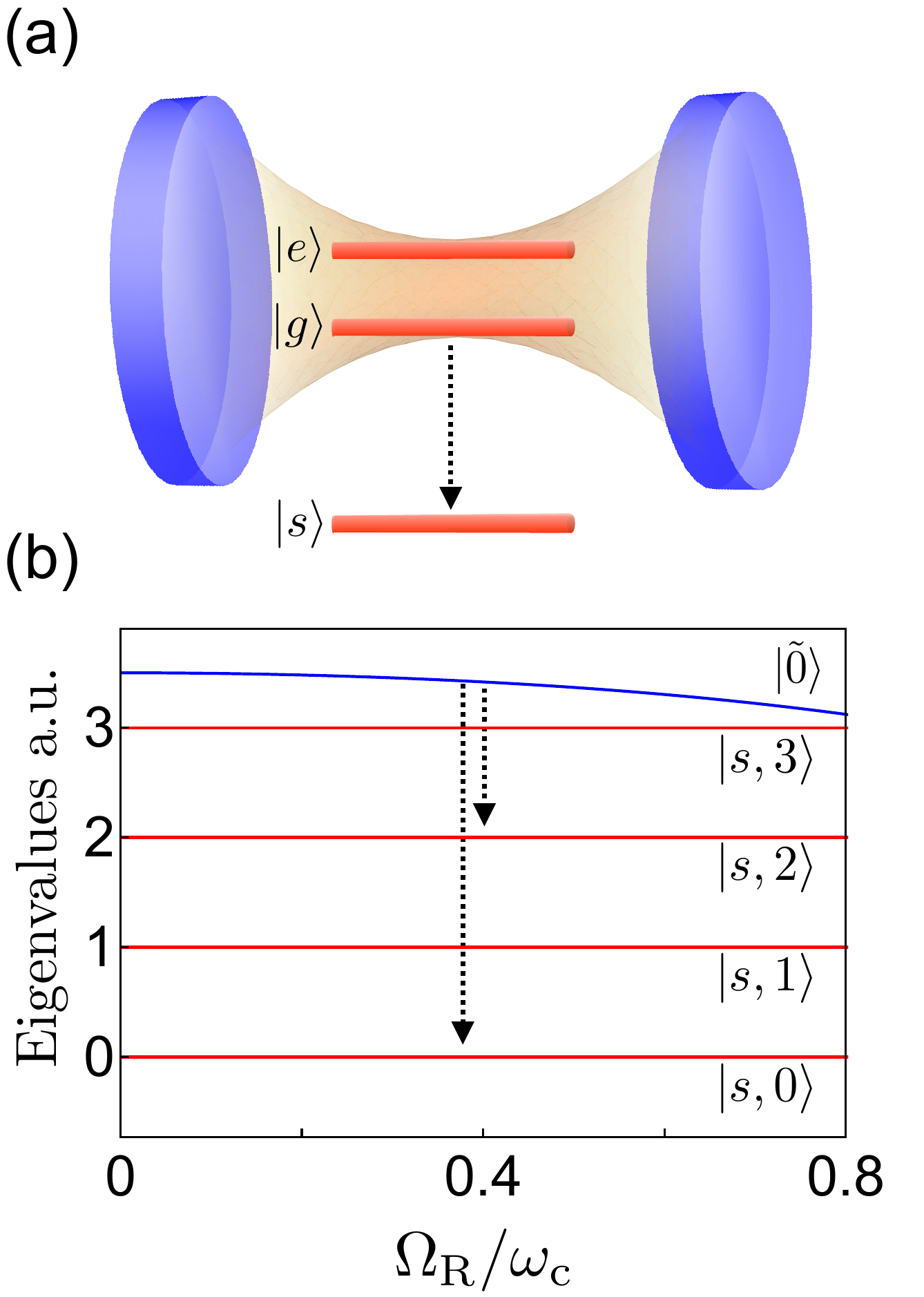}
	\caption{(Color online)
		(a) Sketch of a cavity-embedded three-level system. Only the upper transition $| g \rangle \leftrightarrow | e \rangle$ interacts with the cavity mode. The lowest energy state of the three-level system is $|s \rangle$. (b) Energy spectrum of $\hat H$ as a function of the coupling strength $\Omega_{\rm R}/\omega_{\rm c}$, with $\omega_{gs}=3.5\omega_{\rm c}$ and  $\omega_{eg}=\omega_{\rm c}$. The red horizontal lines represent the non-interacting states $|s,n\rangle$, the blue curve is the lowest-energy atom-cavity dressed state $|\tilde{0}\rangle$. The black arrows indicate the transitions stimulated by the driving pulses.}\label{3lev}
\end{figure}
The system Hamiltonian is
\begin{equation}\label{eq:model}
\hat H = \omega_{\rm c} \hat a^{\dagger}\hat a + \sum_{\alpha= {\rm s, g,e}} \omega_\alpha \hat \sigma_{\alpha \alpha}  + \Omega_{\rm R} ( \hat a + \hat a^{\dag})(\hat \sigma_{\rm eg}+ 
\hat \sigma_{\rm ge})\, ,
\end{equation}
where $\omega_{\alpha}$ ($\alpha={\rm s,g,e}$) are the bare frequencies of the atom-like relevant states, and $\sigma_{\alpha \beta} = |\alpha \rangle\langle \beta|$ describes the transition operators (projection operators if $\alpha= \beta$) involving the levels of the quantum emitter.   The Hamiltonian can be separated as $\hat H = \hat H_{\rm R} + \hat H_{\rm s}$, where $\hat H_{\rm R}$ is the well known Rabi Hamiltonian, Eq.\ (\ref{HR}), and $\hat H_{\rm s} = \omega_{\rm s}\hat \sigma_{\rm s s}$.
As a consequence, the total Hamiltonian is block-diagonal and its eigenstates can be separated into a non-interacting sector $|{\rm s}, n \rangle$, with energy $\omega_{\rm s}+ n \omega_{\rm c}$, where $n$ labels the cavity photon number, and into  dressed atom-cavity states $| \tilde{j} \rangle$, resulting from the diagonalization of the Rabi Hamiltonian. 
We consider the system  initially prepared in the $|\tilde{0}\rangle$ state. Preparation can be accomplished by simply exciting the system initially in the ground state $| s,0 \rangle$ with a $\pi$ pulse of central frequency $\omega_{\tilde 0} - \omega_s$. Then   the qubit is excited by two additional pulses with central frequencies  $\omega_1 =\omega_{\tilde 0} - 2 \omega_{\rm c}$ and $\omega_2 = \omega_{\tilde 0}- \omega_{\rm s}$.
The driving Hamiltonian is
\begin{equation}\label{pulse2}
\hat H_{\rm d} =\left[ {\cal E}_1 (t) \cos (\omega_1 t) +  {\cal E}_2 (t) \cos (\omega_2 t)\right] (\hat \sigma_{gs} + \hat \sigma_{sg})\, ,
\end{equation}
where ${\cal E}_{1,2}(t) = A_{1,2} \exp{[-(t-t_0)^2/(2 \tau^2)]}/(\tau \sqrt{2 \pi})$ describes  Gaussian pulses.
While the transition $|\tilde{0}\rangle \to |s,0\rangle$ is allowed  in the weak-coupling regime or even in the absence of a resonator, the matrix element for the transition $|\tilde{0}\rangle \to |s,2\rangle$ vanishes for a zero coupling rate and is negligible until $\Omega_{\rm R}$ reaches at least  $10 \%$ of $\omega_{\rm c}$. Specifically:
\begin{eqnarray}
\langle s,0 | (\hat \sigma_{gs} + \hat \sigma_{sg}) | \tilde{0} \rangle = c_{g,0}^{\tilde 0}\, , \nonumber \\
\langle s,2 | (\hat \sigma_{gs} + \hat \sigma_{sg}) | \tilde{0} \rangle = c_{g,2}^{\tilde 0}\, .
\end{eqnarray}
In order to obtain a quantum superposition $\cos \phi|s,0 \rangle + \sin \phi | s,2 \rangle$ via the dressed vacuum state $|\tilde{0}\rangle,$ the pulse amplitudes have to satisfy the following relationship: $  A_1 c_{g,0}^{\tilde 0} /   A_2 c_{g,2}^{\tilde 0} = \tan \phi$. In order to obtain large squeezing, we choose the driving amplitude such that $ \tan \phi \approx \sqrt{2}/2$, corresponding to the angle where  squeezing for this superposition state is maximal.
\begin{figure}[ht!]
	\includegraphics[width=80 mm]{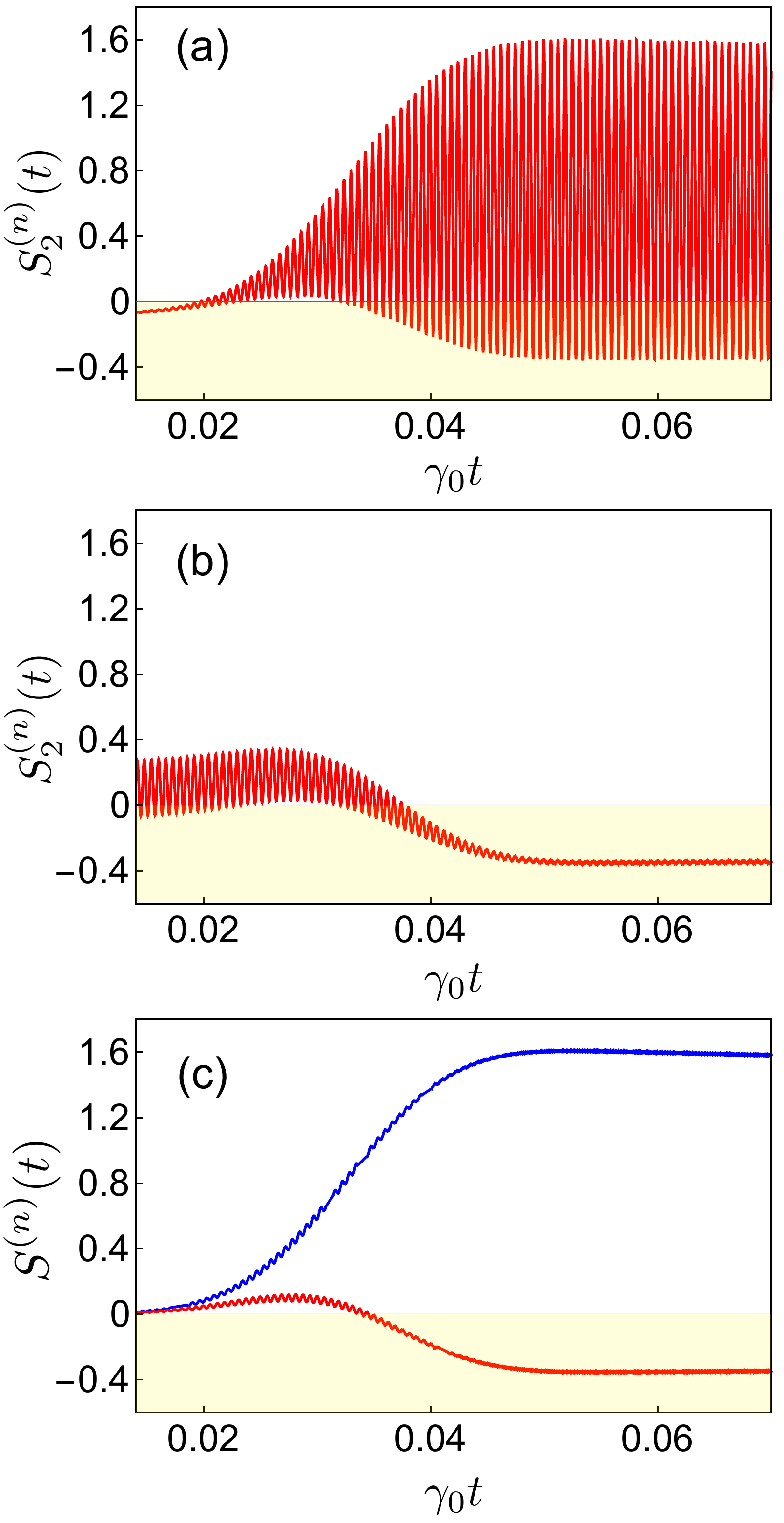}
	\caption{(Color online) Time evolution of the normally-ordered variances $S^{(n)}_i$ for a system initially prepared in the lowest-energy dressed state $|\tilde {0}\rangle$. The qubit is excited by two pulses with central frequencies $\omega_1 =\omega_{\tilde 0} - 2 \omega_{\rm c}$ and $\omega_2 = \omega_{\tilde 0}- \omega_{\rm s}$,   at the time $\gamma_0 t_1=2.6\times 10^{-2} $ and $\gamma_0 t_2=3.8\times 10^{-2} $, respectively, and with amplitude such that $ \tan \phi \approx \sqrt{2}/2$. The damping rates are $\gamma_{\rm c}=\gamma_{\rm eg}=\gamma_{\rm gs}=2\times 10^{-4}\omega_{\rm c}$, and the coupling constant is $\Omega_{\rm R}=0.4\omega_{\rm c}$. Other parameters are the same in Fig.\,4. (a, b) The variances $S^{(n)}_2$ are calculated using standard operators with reference frequency (a) $\Omega=0$ and (b) $\Omega=\omega_{\rm c}$. (c) The variances $S^{(n)}_1$ (blue upper curve) and $S^{(n)}_2$ (red lower curve) are calculated using the correct positive and negative operators for reference frequency $\Omega=\omega_{\rm c}$.}\label{Dyn3lev}
\end{figure}
Figures~5(a) and 5(b) display the time evolution of the variances $S^{(n)}_2$ calculated with the standard operators ($\hat x^+ = \hat a$), while Fig.\,5(c) displays the time evolution of the variances $S^{(n)}_1$ (blue curve) and $S^{(n)}_2$ (red curve) using the correct positive and negative field operators. The behavior of $S^{(n)}_2(t)$ in Fig.\,5(a) starts with a fictitious value less than zero, while in Fig.~5(c) correctly starts from $0$. The variance $S^{(n)}_2$ in Fig.\,5(a) has been calculated by using the reference frequency $\Omega= 0$. Figure 5(b) has been obtained by using $\Omega=\omega_{\rm c}$.
These different choices show that it is not possible to eliminate fast and large-amplitude fictitious oscillations within the standard approach. Figure 5(c) has been obtained  with $\Omega =  \omega_{\rm c}$.

\section{Conclusions}
We have derived a generalized theory of the output field-quadrature measurements and squeezing in cavity-QED systems, valid for arbitrary cavity-atom coupling rates. 
In the USC regime, where the counter-rotating terms cannot be ignored, the standard theory predicts a large amount of squeezing in 
the output field, even when the system is in its ground state. Here we have shown that, in this case, no squeezing can be detected in the output field-quadratures, independently of the system details. 
We have applied our theoretical approach to study the output squeezing produced by an artificial two-level atom embedded in a coherently excited cavity.
We also studied the output field-quadratures from  a cavity interacting in the USC regime with the upper transition of a cascade-type three-level 
system. 
The numerical results  have been compared with 
the standard calculations  of output squeezing. 
The approach proposed here can be directly applied also to resonators displaying  ultrastrong optical nonlinearities \cite{Ridolfo2013}. This work extends the possibilty of predicting and  analyzing output-field correlations when optical resonators interact very strongly with other quantum systems.

\section*{Acknowledgements}

We thank Professor Adam Miranowicz for very useful discussions. This work is partially supported by the RIKEN iTHES Project, the MURI Center for Dynamic Magneto-Optics via the AFOSR award number FA9550-14-1-0040,
the IMPACT program of JST, a Grant-in-Aid for Scientific Research (A), and  from the MPNS
COST Action MP1403 Nanoscale Quantum Optics.

\appendix
\section{MASTER EQUATION}
In the ultrastrong coupling regime, owing to the high ratio $\Omega_{\rm R} /\omega _{\rm c}$, the standard approach fails to correctly describe the dissipation processes and leads to unphysical results as well. In particular, it predicts that even at $T=0$, relaxation would drive the system out of its ground state $|G \rangle$ generating photons in excess to those already present.\\ 
The right procedure that solves such issues consists in taking into account the atom-cavity coupling when deriving the master equation after expressing the Hamiltonian of the system in a basis formed by the eigenstates $|j\rangle$ of the Rabi Hamiltonian $\hat H_{\rm R}$. The dissipation baths are still treated in the Born-Markov approximation. 
Following this procedure it is possible to obtain the master equation in the dressed picture \cite{Blais2011}. 
For a $T=0$ reservoir, one obtains: 

\begin{equation} 
\label{9}
\hat{\dot{\rho }}(t)=-i\left [\hat H_{\rm S},\hat \rho(t) \right ]+\mathcal{L}_{\rm a} \hat \rho(t)+\mathcal{L}_{\rm x}\hat \rho(t)\;.\end{equation}

Here $\mathcal{L}_{\rm a}$ and $\mathcal{L}_{\rm x}$ are the Liouvillian superoperators correctly describing the losses of the system where $\mathcal{L}_{\rm s}\hat \rho(t)=\sum_{j,k>j}\Gamma_{\rm s}^{jk} \mathcal{D}[|j\rangle\langle k|]\hat \rho(t)$ for $s=a,\sigma_{-}$ and $\mathcal{D}[\hat O]\hat\rho=\frac{1}{2}(2 \hat O\hat\rho \hat O^{\dagger}-\hat \rho \hat O^{\dagger}\hat O-\hat O^{\dagger}\hat O \hat \rho)$.  In the limit $\Omega_{\rm R} \rightarrow 0$, standard dissipators are recovered. 

The relaxation rates $\Gamma_{s}^{jk}=2\pi d_{s}(\Delta_{kj} )\alpha_{s}^{2}(\Delta_{kj} )\left | C_{jk}^{s} \right |^{2}$ depend on the density of states of the baths $d_{\rm s}(\Delta_{kj})$ and the system-bath coupling strength $\alpha_{\rm s}(\Delta_{kj} )$ at the respective transition frequency $\Delta_{kj}\equiv\omega_{k} - \omega_{j}$ as well as on the transition coefficients $C_{jk}=\langle j|\hat s+\hat s^{\dagger}|k\rangle \; (\hat s=\hat a,\hat \sigma_{-} )$. These relaxation coefficients can be interpreted as the full width at half maximum of each $|k\rangle\rightarrow |j\rangle$ transition. In the Born-Markov approximation the density of states of the baths can be considered a slowly varying function of the transition frequencies, so that we can safely assume it to be constant as well as the coupling strength.

\bibliography{squeez}

\end{document}